\documentclass[10pt,letterpaper,compsoc,conference]{iiswc25}

\usepackage{amsmath,amssymb,amsfonts}
\usepackage{algorithmic}
\usepackage{graphicx}
\usepackage[dvipsnames]{xcolor}
\usepackage[final]{microtype}
\usepackage[italic]{mathastext}
\usepackage{libertine}
\usepackage[T1]{fontenc}
\usepackage{textcomp}
\usepackage[varqu,varl]{zi4}
\usepackage[all]{nowidow}
\usepackage[auth-lg,affil-it]{authblk}
\usepackage[keeplastbox]{flushend}
\usepackage{fancyhdr}

\usepackage[bibstyle=ieee,citestyle=numeric-comp,sorting=nyt,maxbibnames=99]{biblatex}
\usepackage{comment}
\usepackage{makecell}
\usepackage{listings}
\usepackage{enumitem}
\usepackage{siunitx}

\usepackage{adjustbox}
\usepackage{array}

\usepackage{multirow}
\usepackage{booktabs}

\usepackage[hidelinks]{hyperref}

\newcolumntype{R}[2]{%
    >{\adjustbox{angle=#1,lap=\width-(#2)}\bgroup}%
    l%
    <{\egroup}%
}

\DeclareSIUnit{\us}{\SIUnitSymbolMicro s}

\fancypagestyle{firstpage}{
  \fancyhf{}
  
  \fancyhead[C]{\vspace{10pt}\normalsize{%
      Pre-Print for Publication} \\\vspace{-25pt}} 
  \fancyfoot[C]{\thepage}
}

\begin{document}


\title{Dissecting CPU-GPU Unified Physical Memory \\on AMD MI300A APUs}


\renewcommand\Authsep{\qquad}
\renewcommand\Authand{\qquad}
\renewcommand\Authands{\qquad}


\author[1]{Jacob Wahlgren}
\author[1]{Gabin Schieffer}
\author[1]{Ruimin Shi}
\author[2]{Edgar A. León}
\author[2]{\\Roger Pearce}
\author[2]{Maya Gokhale}
\author[1]{Ivy Peng}
\affil[1]{KTH Royal Institute of Technology, Sweden}
\affil[2]{Lawrence Livermore National Laboratory, USA}

\maketitle
\thispagestyle{firstpage}
\pagestyle{plain}


\begin{abstract}
Discrete GPUs are a cornerstone of HPC and data center systems, requiring management of separate CPU and GPU memory spaces. Unified Virtual Memory (UVM) has been proposed to ease the burden of memory management; however, at a high cost in performance. The recent introduction of AMD’s MI300A Accelerated Processing Units (APUs)—as deployed in the El Capitan supercomputer—enables HPC systems featuring integrated CPU and GPU with Unified Physical Memory (UPM) for the first time. This work presents the first comprehensive characterization of the UPM architecture on MI300A. We first analyze the UPM system properties, including memory latency, bandwidth, and coherence overhead. We then assess the efficiency of the system software in memory allocation, page fault handling, TLB management, and Infinity Cache utilization. We propose a set of porting strategies for transforming applications for the UPM architecture and evaluate six applications on the MI300A APU. Our results show that applications on UPM using the unified memory model can match or outperform those in the explicitly managed model---while reducing memory costs by up to 44\%. 
\end{abstract}

\section{Introduction}
\label{sec:intro}
GPUs are a critical component of leadership clusters and high-performance computing (HPC) systems for their massive parallelism and high computing power. Today, most top supercomputers are equipped with discrete GPUs~\cite{top500}, with separate memory spaces for the CPU and the GPU. The memory speed lags behind as the computing speed continually improves, causing efficient data access to become increasingly critical for exploiting the full potential of emerging systems~\cite{dongarra-turing-award}. Applications that require frequent data movement between the CPU and GPU memories suffer from degraded performance and increased energy usage. In the past decade, extensive works have been proposed for optimizing memory access and reducing data movements between CPU and GPU~\cite{chen2014porple,agarwal2015page,shen2018cudaadvisor,lin2023drgpum}.

To improve developer productivity, Nvidia introduced Unified Virtual Memory (UVM), enabling a unified memory programming model without the programmer explicitly managing data movement between CPU and GPU memory spaces~\cite{landaverde2014investigation, chien2019performance, allen2021demystifying}. UVM greatly simplifies code development by relying on the runtime software to transparently migrate data between CPU and GPU memories. However, as software solution, it also introduces significant performance penalties due to page faulting and page migrations~\cite{ganguly2019interplay, allen2021demystifying, allen2021depth}. One study found that performance often degrades by 2--3$\times$ and sometimes as much as 14$\times$ compared with traditional explicit memory management~\cite{chien2019performance}.
Many works have proposed runtime and system optimizations, including batching, prefetching, preeviction, and migration mechanisms to enhance the performance of UVM~\cite{landaverde2014investigation,chien2019performance,allen2021demystifying,allen2021depth,cooper2024shared}. Vendors have also explored architectures with more tightly integrated CPU and GPU memory, such as the Grace Hopper Superchip from Nvidia and the MI250X from AMD. Nonetheless, performance remains workload-dependent, and applications using the unified memory model enabled by UVM often result in suboptimal performance compared to the explicitly managed memory model.

In contrast to UVM, Unified Physical Memory (UPM) enables the unified memory programming model with hardware support. In a UPM system, a single physical memory space is shared by the CPU and GPU. AMD has recently introduced the first UPM architecture for HPC and Data Centers -- the MI300A APU~\cite{smith2024realizing}, which is used to implement El Capitan, the No.~1 supercomputer on the Top 500 list~\cite{top500}. UPM could fundamentally eliminate the performance overhead in previous software-based unified memory solutions like UVM. However, nearly all existing HPC applications are using the explicitly managed model due to its superior performance compared to the unified memory model using UVM. With the emergence of UPM, this work aims to answer the open question of whether the unified memory model can now compete with the performance of explicit management. 

In this work, we provide a timely full-stack characterization study of the UPM architecture on the AMD MI300A APU, including the system properties and system software support, as well as application-level performance.
Our characterization methodology includes standard benchmarks and custom benchmarks specifically designed for the UPM architecture, as well as detailed insights from profiling tools, and a study of six HPC workloads from the Rodinia suite~\cite{che2009rodinia}. We highlight differences between memory allocators regarding performance and overhead, and analyze TLB management and Infinity Cache utilization on the UPM on MI300A APU. We identify a set of porting strategies for transitioning existing codes from the explicit model to the unified memory model. By analyzing six applications on MI300A APU, we find that UPM enables the unified memory model to have performance on par with the explicitly managed model, while additionally reducing memory cost by up to 44\%. This impressive memory saving on the UPM system enables much larger problems on one APU within a smaller envelope compared to traditional discrete GPUs.
In summary, we made the following main contributions in this work:

\begin{itemize}[leftmargin=*]
    \item We provide an in-depth characterization of the unified physical memory architecture on the MI300A, including latency, bandwidth, and coherence overhead.
    \item We quantify the efficacy of system software support for memory management on UPM on MI300A, including memory allocation, page fault handling, TLB management, and Infinity Cache utilization.
    \item We transform six HPC workloads into the unified memory model and compare their performance with that from the explicitly managed model on UPM.
    \item Our results highlight that UPM enables the performance of the unified memory model to be on par with the explicitly managed model, while saving memory cost by up to 44\%, when applying our porting strategies.
\end{itemize}

\section{Unified CPU--GPU Memory}

GPU memory is usually explicitly managed as a separate memory space because the CPU and GPU have physically separated memories, e.g. with the GPU connected through PCIe as a peripheral device. This discrete GPU architecture naturally leads to the popularity of \textit{explicitly managed memory model}, as exemplified in Listing~\ref{code1}, where separate memory allocations, e.g. via the \verb|malloc| and \verb|hipMalloc| allocators, are needed in the CPU and GPU memories, and data is copied between them explicitly via e.g. \verb|hipMemcpy|. As a result, data is duplicated in both the CPU and GPU memories. Nevertheless, this explicit model is the most commonly used programming model in today's GPU applications due to its high performance.

To improve programming productivity, the \textit{unified memory programming model}, as exemplified by Listing~\ref{code2}, was introduced to support unified memory allocations in CPU and GPU memories, and avoid the need for explicitly initiated data movement. The unified memory model can be implemented by either software-based solutions such as UVM (e.g., via using \verb|hipMallocManaged| allocator and implicit data movement triggered by page faults), or hardware-based solutions such as UPM, i.e., a single physical memory shared by CPU and GPU, eliminating the need for data movement.

\subsection{Unified Virtual Memory}
UVM enables the unified programming model by providing the illusion of a single coherent CPU--GPU memory by leveraging transparent page fault handling and page migration between CPU and GPU memories. Nvidia introduced UVM in CUDA~6 with the \verb|cudaMallocManaged()| allocator.  Since the Pascal GPU architecture, there is a dedicated hardware unit for page translation and migration that enables accessed pages in \verb|cudaMallocManaged|-allocated memory regions to be migrated on-demand. However, current UVM-based unified memory model can cause significant performance impact on applications due to page faults and migration costs~\cite{landaverde2014investigation,chien2019performance,allen2021demystifying,cooper2024shared}. To mitigate performance overhead from page fault handling and page migration, several works propose optimizations for page prefetching and pre-eviction~\cite{chien2019performance,allen2021demystifying,}.

Recently, more tightly connected CPU--GPU memory is supported by cache-coherent interconnects, such as the NVLink-C2C in Nvidia's Grace Hopper Superchip and Infinity Fabric on AMD's MI250X. These architectures support high-bandwidth low-latency data transfer between CPU and GPU memory, mitigating the page fault and migration overhead in traditional UVM by enabling the GPU to directly access CPU memory at cacheline granularity. However, the unified memory programming model on Grace Hopper and MI250X still needs to manage separate physical CPU and GPU memories.

A benefit of UVM over UPM is that it enables over-committing memory on the GPU by utilizing host memory.

\subsection{Unified Physical Memory in MI300A}
In UPM architectures, CPUs and GPUs are integrated into the same die and share one physical memory. Such integrated CPU--GPU units are known as Accelerated Processing Units (APUs) on AMD systems. The AMD MI300A was recently released as the first APU targeting HPC systems. The current No. 1 HPC system on the Top500 list, the El Capitan supercomputer, features 44,544 MI300A APUs. UPM simplifies the system architecture as separate CPU chips and memory are not needed. Further, the architecture natively supports the unified memory programming model, and the high overhead of software management needed by UVM can be completely eliminated on the physically unified hardware. Finally, CPU--GPU data transfers, which are often the bottleneck in existing GPU codes, are no longer needed on UPM.

\begin{figure}[bt]
    \centering
    \includegraphics[width=\linewidth]{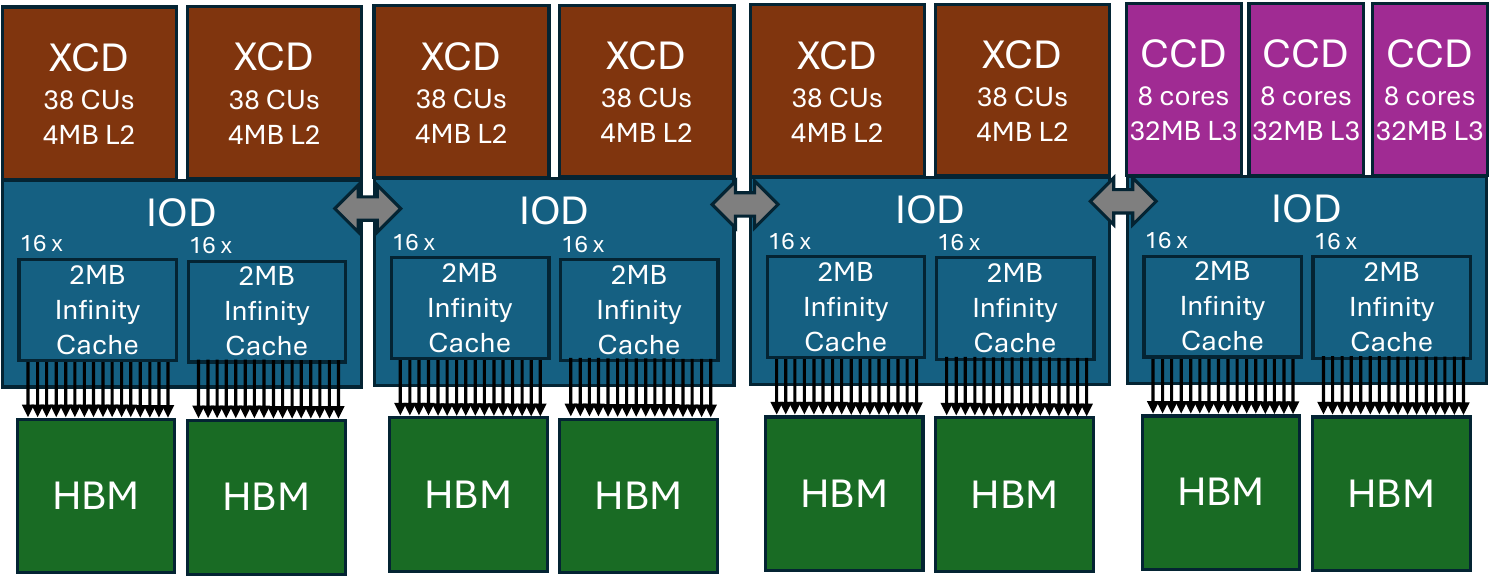}
    \caption{An overview of the chiplet-based MI300A APU architecture including six XCD (GPU) and three CCD (CPU).}
    \label{fig:arch}
\end{figure}

In this work, we focus on the UPM on the MI300A APU as it represents state-of-the-art HPC systems. The MI300A APU is enabled by chiplet integration and is based on the AMD CDNA 3 architecture~\cite{cdna3-whitepaper,smith2024realizing}. As illustrated in Fig.~\ref{fig:arch}, the GPU part consists of six accelerator complex dies (XCDs) while the CPU part consists of three CPU complex dies (CCDs). The six XCDs are presented as a single device to the user in the standard configuration. Every two XCDs or three CCDs share an IO die (IOD). Four IODs on one APU implement cross-die communication and the HBM3 interface to eight memory stacks. Each memory stack has 16 memory channels and 16~GiB capacity. The AMD Infinity Fabric interconnects CCD and XCD chiplets and routes memory requests to memory channels. In total, each MI300A APU has 128~GiB HBM3 memory and a peak theoretical bandwidth of 5.3~TB/s.

\begin{figure}[bt]
\begin{minipage}[b]{0.475\linewidth}
\begin{lstlisting}[language=C,caption={Explicit model.},
captionpos=b,label=code1]
float *h = cpu_alloc(n);
float *d = gpu_alloc(n);
init_on_cpu(h);
copy_to_gpu(d, h, n);
gpu_kernel<<<...>>>(d);
copy_to_cpu(h, d, n);
\end{lstlisting}
\end{minipage}%
\hfill
\begin{minipage}[b]{0.475\linewidth}
\begin{lstlisting}[language=C,caption={Unified model.},
captionpos=b,label=code2]
float *u = uni_alloc(n);
init_on_cpu(u);
gpu_kernel<<<...>>>(u);
gpu_synchronize();
\end{lstlisting}
\end{minipage}
\end{figure}

The cache hierarchy consists of two levels in the GPU, three levels in the CPU, and a 256~MiB \textit{Infinity Cache}. Atomic operations in the GPU are implemented with dedicated atomic units located in the shared L2 cache~\cite{mi300-isa}, while the CPU implements atomics by taking exclusive ownership of the data in the private L1 cache~\cite{inside-mi300a}. The Infinity Cache is a memory-side cache shared between the CPU and GPU, and is a new feature of the AMD CDNA 3 architecture, aiming to increase cache bandwidth and reduce off-chip memory accesses. The peak bandwidth from the Infinity Cache can reach 17.2~TB/s, approximately $3\times$ the main memory bandwidth. The Infinity Cache does not participate in coherency and thus does not need to absorb or handle any snoop traffic, significantly improving efficiency and reducing the latency of snooping from other cache levels. It can also hold nominally uncacheable memory such as I/O buffers.

The runtime API for programming AMD GPUs is called HIP (Heterogeneous-compute Interface for Portability). It provides C++ interfaces for writing and launching GPU kernels, managing GPU memory, synchronizing CPU and GPU, etc.

\subsection{Memory Allocation on MI300A}
Two page tables are used for managing address translation on MI300A---a system page table on the CPU and a GPU page table on the GPU. Unlike Nvidia's Grace Hopper, where the GPU can access both page tables, the GPU on MI300A can only access its own page table. Thus, page table entries (PTEs) must be propagated from the system page table to the GPU table to enable GPU access. The two copies are kept in sync using the Linux kernel's heterogeneous memory management (HMM) subsystem.

Table~\ref{tab:allocator} lists memory allocators on MI300A and classifies their physical memory allocation as either on-demand or up-front. \textit{Up-front} allocators allocate all physical pages immediately when the allocator is called while \textit{on-demand} allocators defer physical allocation until the first touch, relying on virtual memory management with page faults.

First, \verb|malloc| is the standard libc function for allocating (host) memory, and serves as a representative for any standard memory allocator, including e.g. C++'s \verb|new|. \verb|hipHostRegister| is used to make host memory (e.g. from \verb|malloc|) accessible on GPUs by locking the pages and mapping them in the GPU page table. \verb|hipHostMalloc| directly allocates GPU-accessible page-locked memory. \verb|hipMalloc| is the standard GPU memory allocator. Finally, \verb|hipMallocManaged| is traditionally called to allocate UVM buffers, which are accessible from both CPU and GPU through migration (although no migration is used on UPM).

By default, the MI300A GPU does not resolve page faults, i.e., it cannot access on-demand mapped pages. To enable the GPU to resolve page faults, AMD GPUs feature a mechanism known as XNACK ("non-acknowledgment") in the TLB, which enables page faults to be replayed~\cite{bertolli2024performance}. With XNACK, when a page fault occurs, the TLB waits for the PTE to be updated by the fault handler before retrying the memory access, thus allowing access to on-demand mapped memory.

\begin{table}[tb]
    \centering
    \caption{Memory Allocators on MI300A}
    \label{tab:allocator}

\adjustbox{max width=\linewidth}{%
    \begin{tabular}{lccl}
    \toprule
    \textbf{Allocator}
    & \makecell{\textbf{GPU} \\ \textbf{Access}}
    & \makecell{\textbf{CPU} \\ \textbf{Access}}
    & \makecell{\textbf{Physical} \\ \textbf{Allocation}} \\
    \midrule
     \verb|malloc|                   &            & \checkmark & On-demand\\
     \verb|malloc (XNACK=1)|             & \checkmark & \checkmark & On-demand\\
     \verb|malloc + hipHostRegister| & \checkmark & \checkmark & Up-front \\
     \verb|hipMalloc|                & \checkmark & \checkmark & Up-front\\
     \verb|hipHostMalloc|            & \checkmark & \checkmark & Up-front\\
     \verb|hipMallocManaged|         & \checkmark & \checkmark & Up-front \\
     \verb|hipMallocManaged (XNACK=1)|   & \checkmark & \checkmark & On-demand \\
    \bottomrule
    \end{tabular}}
\end{table}

\section{Characterization Methodology}

\begin{table}[tb]
    \centering
    \caption{Overview of Experimental Method}
    \label{tab:method}
    \adjustbox{max width=\linewidth}{%
    \begin{tabular}{lll}
        \toprule
        \multirow{6}{*}{\textbf{Benchmarks}}
        & Memory latency & multichase~\cite{google-multichase,fusco2024understanding} \\
        & Memory bandwidth & STREAM~\cite{stream,hip-stream} \\
        & Legacy transfer & hip-bandwidth~\cite{rocm-examples} \\
        & Coherence overhead & Custom \\
        & Allocation speed & Custom \\
        & Page fault overhead & Custom \\
        \midrule
        \multirow{4}{*}{\textbf{Profiling tools}}
        & Memory usage & libnuma \\
        & GPU fragment size & rocprofv3 \\
        & CPU allocation size & perf \\
        & Code generation & hipcc -save-temps \\
        \midrule
        \multirow{6}{*}{\textbf{HPC workloads}}
        & \multirow{6}{*}{Rodinia suite~\cite{che2009rodinia}}
        & backprop \\
        && dwt2d \\
        && heartwall \\
        && hotspot \\
        && nn \\
        && srad\_v1 \\
        \bottomrule
    \end{tabular}}
\end{table}

We summarize our characterization method, including benchmarks, profiling tools, and HPC applications, in Table~\ref{tab:method}. All of our benchmarks are open-source and available at \url{https://github.com/KTH-ScaLab/mi300a-benchmarks}.

We run experiments on a testbed equipped with four AMD MI300A APUs per node. Each APU has 228 GPU compute units (CUs) and 24 CPU cores, and 128~GiB HBM3 memory. The software environment includes Cray Programming Environment~24.11 and ROCm~6.3.1. We use \verb|numactl| and \verb|HIP_VISIBLE_DEVICES| to bind experiments to a single APU on MI300A.

\subsection{Benchmarks}
\textbf{Memory Latency.} We use a pointer-chasing benchmark adapted from Google's multichase~\cite{google-multichase}. The GPU version is based on a CUDA port~\cite{fusco2024understanding}, which we modified to support HIP. We added the ability to use different memory allocators. The benchmark uses a persistent kernel that periodically increments an atomic counter, which measures the memory access time at a granularity of 200 accesses over 0.5~s per iteration from a CPU thread. We set the cache flush size to 256~MiB, the number of sample iterations to 10, and varied the buffer size from 1~KiB to 4~GiB.

\textbf{Memory Bandwidth.}
We used a modified STREAM benchmark to measure the achievable memory bandwidth using the TRIAD kernel. For CPU, we used the standard STREAM implementation~\cite{stream}, and for GPU we used hip-stream~\cite{hip-stream,cuda-stream}. We modified the benchmarks to support different memory allocators and data initialization on either CPU or GPU.
The CPU benchmark used \verb|OMP_PROC_BIND=true| and various number of threads from 1 to 24, selecting the best results. The CPU array size was 610~MiB, and the GPU array size was 256~MiB.

\textbf{Legacy CPU--GPU Data Transfer.}
Many existing applications are written assuming separate memory spaces for CPU and GPU. These legacy applications can run on MI300A. However, they may incur unnecessary data transfer overhead between "host memory" and "device memory", which no longer exist on UPM.
We evaluate the cost of legacy data transfer by measuring the bandwidth of \verb|hipMemcpy| with the \verb|hip_bandwidth| benchmark~\cite{rocm-examples}.

\textbf{Coherence Overhead.}
Many lock-free algorithms rely on high-performance atomics to resolve data races. However, parallel atomics imply coherence overhead that increases with the level of contention. The shared physical memory between CPU and GPU on UPM could further exacerbate contention and coherence overhead as data needs to be available to both the CPU and GPU.

We designed a benchmark to measure the performance of atomic operations and the coherence overhead when CPU and GPU operate on the same data structure. The benchmark computes a parallel histogram, where an array is initialized to zero, and then randomly selected elements are incremented in a loop using atomic addition. Both CPU threads and GPU threads can be used to perform this update. The throughput is measured similarly to the multichase benchmark by periodically reading an update counter from a separate CPU thread.

The CPU kernel uses \verb|std::minstd_rand| uniform distribution to generate random numbers and is launched using \verb|std::thread|. The compiler intrinsic \verb|__atomic_fetch_add()| is used to implement atomic increment. The GPU kernel uses 64 threads per block and generates random numbers using XORWOW generator in the rocRAND library. Atomic increment on the GPU is implemented with \verb|atomicAdd_system()|. (Note that the function \texttt{atomicAdd\_system} is documented as "system scope" while \texttt{atomicAdd} is documented as "device scope". They determine the sc1 bit in the generated instruction. However, we did not observe any difference between the two in performance or correctness.)

We performed experiments using four array sizes: 1, 1K, 1M, 1G (i.e. $2^0, 2^{10}, 2^{20}, 2^{30}$). The 1 and 1K cases fit in L1 cache, 1M fits in L2 cache, and 1G does not fit in any cache. The array contains either integer (UINT64) or floating point (FP64) elements.

\textbf{Allocation Speed.}
Understanding the performance of different memory allocators is important for applications like adaptive mesh refinement~\cite{berger1989local} and Lagrangian hydrodynamics~\cite{LULESH:spec}, which require allocating and deallocating memory dynamically at runtime. We design a benchmark consisting of two loops. The first loop allocates N~chunks of memory of size~M, while the second loop frees the chunks. We used 10 warmup iterations and set N to 100. We measure the loops using a CPU timer. There is no need for explicit device synchronization since all the allocators are inherently synchronous. We allocate 2~B to 1~GiB of data. This benchmark excludes the time to touch the allocated memory, which is studied separately in the next section on page faults.

\textbf{Page Fault Overhead.}
On-demand memory allocators offer low latency, but come with a cost of page faults at runtime on the first touch of each page. We design a benchmark for quantifying the latency and throughput for handling different types of page faults on MI300A.
The page fault overhead is the difference in runtime between accessing an already mapped page and accessing an unmapped page (causing a page fault).
Our benchmark issues a single load to each page, and we measure the page fault time as the difference between running it on a newly allocated array (the faulting version) and a pre-faulted array (the non-faulting version) with a CPU timer. On the GPU, we launch a kernel to access the pages and measure the time from kernel submission to completed device sync.
For memory allocation, we use \verb|mmap| to ensure that each test is independent (\verb|malloc| may allocate a larger chunk of address space, which is faulted in batch). The non-faulting baseline on the GPU is implemented with \verb|hipHostRegister|, and on the CPU with \verb|mlock|. 
We varied parameters such as how many pages are accessed concurrently and investigated four scenarios. In \textit{GPU Major}, on-demand memory is allocated and directly accessed by the GPU. In \textit{GPU Minor}, the allocated memory is touched by the CPU before measuring the fault overhead on the GPU. \textit{1CPU} uses a single CPU core for memory access, while \textit{12CPU} uses 12 cores. The benchmark consists of 10 warm-up iterations followed by 100 timed iterations.

\subsection{Profiling Tools}
\label{sec:memory_usage}

\textbf{Memory Usage.}
No single memory profiling interface can provide a complete picture of memory allocations on MI300A yet. Linux provides system-level memory usage via \verb|/proc/meminfo|, and the libnuma interface reports the free memory per NUMA node, i.e., at the APU level. As expected, both reflect allocations by up-front allocators
immediately, and on-demand allocators
after the first touch. The HIP interface \verb|hipMemGetInfo| and the \verb|rocm-smi| command report free memory "on the device", i.e. at the APU level. However, they only capture allocations by \verb|hipMalloc|. Finally, process-level memory usage can be obtained from the VmRss field in \verb|/proc/pid/status| or the Rss field in \verb|/proc/pid/smaps_rollup| (as displayed in the top command). However, they do not capture allocations by \verb|hipMalloc|.
We choose to profile peak memory usage by sampling from libnuma.

\textbf{GPU Adaptive Fragment Size.}
AMD's GPU page table supports an adaptive scheme that uses \textit{fragments} for improving TLB reach. A fragment is a virtually and physically contiguous range of pages with identical flags. The GPU L1 TLB can store a single entry for a whole fragment, greatly increasing its reach~\cite{amdgpu-fragment-source}. A larger reach means fewer TLB misses and better performance. Each PTE has a 5-bit fragment field, theoretically supporting sizes from a single page (4~KiB) to $2^{31}$ pages (8~TiB). The amdgpu driver sets the fragment field opportunistically by scanning for maximal contiguous page ranges in the fault handler.

The size of fragments in the page table cannot be read directly from userspace. Instead, we use the number of GPU TLB misses as a proxy metric. Using the rocprofv3 GPU profiler, we measure the \verb|TCP_UTCL1_TRANSLATION_MISS_sum| counter to track the number of TLB misses in the GPU. We compare the number of misses in the TRIAD kernel of the GPU STREAM benchmark using different allocators to understand their interaction with memory fragments.

\textbf{CPU Allocation Granularity.}
The memory allocation granularity is impacted by the used allocator and whether CPU or GPU performs first-touch. On the CPU, the number of page faults and TLB misses can imply the granularity. We track these metrics in the CPU STREAM benchmark using \verb|perf stat|.

\textbf{Code Generation.}
To understand which CPU and GPU instructions are generated by the compiler, we use the \verb|-save-temps| flag to the hipcc/clang compiler to output assembly code. In particular, this enables us to understand how atomic operations are implemented.

\subsection{Porting Strategies for Unified Memory}
\label{sec:strat}
This section outlines potential challenges arising from porting codes in the explicit model to the unified memory model and their respective porting strategies. We illustrate each challenge with a simplified example code snippet.

\textbf{Concurrent CPU--GPU Access.}
When a data structure is accessed by CPU and GPU concurrently, simply merging them into a single buffer could result in data race. Without changing the algorithm or imposing synchronization, double buffering could be a preferred solution, i.e. swapping the buffers in each iteration instead of copying.
\begin{lstlisting}
// gpu_kernel overlaps with next cpu_function
for (i = 0; i < n; i++) {
    cpu_function(h_tmp, h_input[i]);
    copy_to_gpu(d_tmp, h_tmp);
    gpu_kernel<<<...>>>(d_tmp, d_sum); }
\end{lstlisting}

\textbf{Memory Usage Consideration.}
Some applications adapt their buffering scheme based on free memory. Existing codes access memory usage counters to determine the amount of free memory capacity. However, as explained in Section~\ref{sec:memory_usage}, previous interfaces for querying memory usage may be inaccurate to reflect all types of memory allocations on UPM. Thus, such applications must change to reliable memory usage counters, such as meminfo or libnuma. Moreover, they must adapt their calculation of free space to consider all types of memory allocations.
\begin{lstlisting}
n = gpu_free_memory() / sizeof(element);
h_array = cpu_alloc(n * sizeof(element));
d_array = gpu_alloc(n * sizeof(element));
\end{lstlisting}

\textbf{Partial Memory Transfer.}
Partial memory transfers arise from situations where only a partial range of a memory buffer are copied between corresponding CPU and GPU buffers. They are often used in a pipeline to overlap data movement with computation, which may become unnecessary in the unified memory model.
\begin{lstlisting}
for (i = 0; i < n; i += chunk_size) {
    cpu_function(h_data+i, chunk_size);
    copy_to_gpu(d_data+i, h_data+i, chunk_size);
    gpu_kernel<<<...>>>(d_data+i, chunk_size); }
\end{lstlisting}

\textbf{Stack Variables.}
While UPM enables the GPU to access the host stack, the asynchronous execution model makes it challenging to analyze the lifetime of stack variables from the GPUs perspective. The host function cannot return until the GPU kernel using the host variable has completed.
\begin{lstlisting}
x = cpu_function();
copy_to_gpu(d_x, x);
gpu_kernel<<<...>>>(d_x, d_sum);
\end{lstlisting}

\textbf{Static Variables.}
Even with UPM, static host memory cannot be accessed from GPU code and vice versa due to linker limitations. The options for unifying static variables are using \textit{managed variables} or modifying the code to use dynamic memory allocation (e.g., \verb|hipMalloc|) instead. The \verb|__managed__| storage specifier is a CUDA/HIP language extension enabling unified variables similar to \verb|hipMallocManaged|. However, it comes with a performance penalty (as we will show in Section~\ref{sec:bw}). On the other hand, using dynamic memory allocation requires restructuring the code.
\begin{lstlisting}
float h_data[100];
__device__ float d_data[100];
\end{lstlisting}

\textbf{Hidden Allocator.}
With libraries that allocate memory on behalf of the user (e.g. C++ containers), it can be challenging to create a high-performance unified allocation. Either a lower-performance allocator will be used (e.g. the default in C++), or the developer has to use more complex APIs or modify the library source code.
\begin{lstlisting}
std::vector h_data;
while (more)
    h_data.push_back(cpu_function());
d_data = gpu_alloc(h_data.size());
copy_to_gpu(d_data, h_data.data());
gpu_kernel<<<...>>>(d_data, h_data.size());
\end{lstlisting}

\subsection{Programming Model Comparison}
We compare the traditional explicit programming model to the UPM-enabled unified memory model on MI300A. We select six applications (shown in Table~\ref{tab:method}) covering a diverse set of coding practices from Rodinia, a widely used suite of GPU-accelerated HPC applications written in CUDA and OpenCL~\cite{che2009rodinia}. For each application, we create two variants. The first variant using the explicit model (corresponding to Listing~\ref{code1}) is a baseline version that ports the original CUDA code to HIP using hipify-perl with minor manual adjustments. The second variant uses the unified memory model (corresponding to Listing~\ref{code2}) by replacing duplicated CPU and GPU allocations with a single unified allocation.

We also modified the input problems to increase the memory usage and runtime. The baseline version uses from 487~MiB memory in heartwall to 43~GiB memory in nn. We use \verb|/usr/bin/time| to measure the total execution time, and inserted timers to measure the time of the main compute phase. The total execution time ranges from 5.23~s in dwt2d to 109~s in nn.

\section{Memory System Characterization}
\label{sec:char}
In this section, we provide a characterization of the UPM system properties.

\subsection{Memory Latency}
\label{sec:latency}
\begin{figure}
    \centering
    \includegraphics[width=\linewidth]{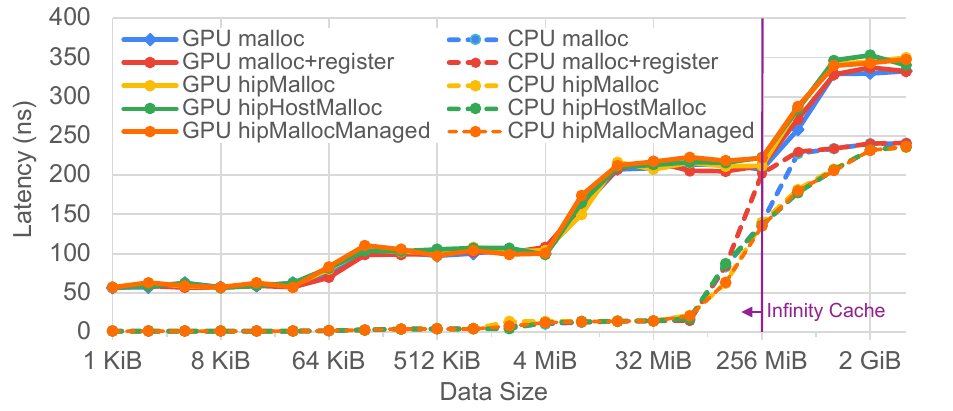}
    \caption{Memory latency on GPU (solid lines) and CPU (dashed lines) with different allocators (semi-log).}
    \label{fig:latency}
\end{figure}

The latency results are shown in Fig.~\ref{fig:latency}. GPU memory accesses reveal three cache levels -- 57~ns at 1~KiB (in L1), 100-108~ns at 1~MiB (in L2), 205-218~ns at 128~MiB (Infinity Cache), and finally 333-350~ns at 4~GiB (in HBM). The CPU memory latency is lower than GPU memory latency. At 1~KiB (in L1), the CPU memory latency is only 1~ns, while at 4~GiB (in HBM), the CPU memory latency is 236-241~ns. For latency-bound tasks, the CPU has a significant advantage over the GPU. The relative difference is especially apparent for data that fits in the CPU L3 cache (96~MiB), which is missing in the GPU.

While GPU memory latency on MI300A is insensitive to the allocator in use, CPU memory latency is not. On the CPU, all allocators eventually plateau at 240~ns latency around 2~GiB. However, between L3 (96~MiB), Infinity Cache (256~MiB) and this plateau point, there is a distinction between the allocators. 
With 256~MiB size, the full dataset fits in IC, and at 512~MiB size, half of the accesses should hit IC. Given that, we would expect the CPU latency at 256-512~MiB to be significantly lower than 240~ns, which is observed with HIP allocators, which increase gradually. However, at 512~MiB, \verb|malloc| and \verb|malloc+register| already result in a latency of 230~ns. This suggests that \verb|malloc| on the CPU cannot leverage the full power of the IC (we explore this further in Section~\ref{sec:ic-util}).

\subsection{Memory Bandwidth}
\label{sec:bw}

\begin{figure}[bt]
    \centering
    \includegraphics[width=\linewidth]{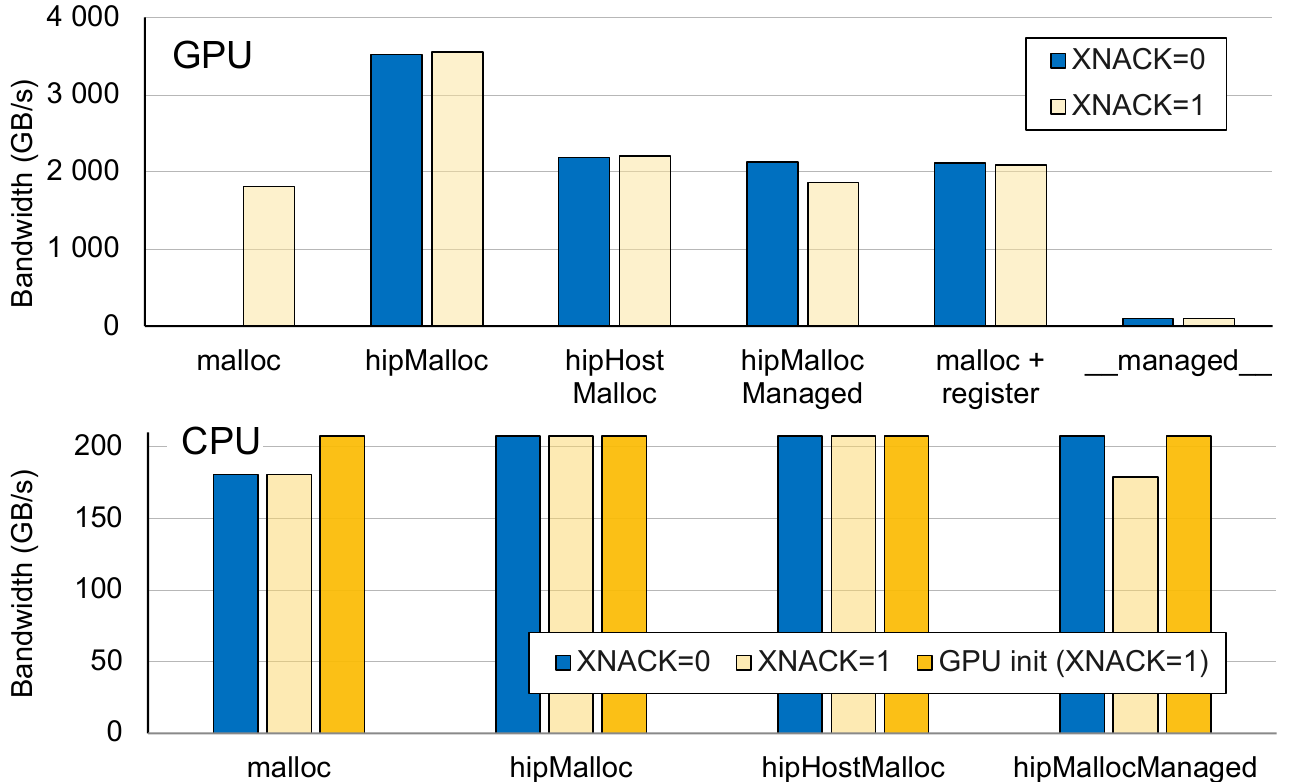}
    \caption{The maximum measured memory bandwidth obtained from GPU (top) and CPU (bottom) using different allocators.}
    \label{fig:stream_cpu_gpu}
\end{figure}

The memory bandwidth results are shown in Fig.~\ref{fig:stream_cpu_gpu}. The GPU memory bandwidth is independent of whether the memory is first touched by the CPU or the GPU. The best GPU bandwidth is achieved with \verb|hipMalloc| at 3.5-3.6~TB/s, while \verb|hipHostMalloc|, \verb|hipMallocManaged| (xnack=0), and \verb|malloc+hipHostRegister| give 2.1-2.2~TB/s. The on-demand allocators \verb|malloc| and \verb|hipMallocManaged| (xnack=1) give the worst performance of the dynamic allocators at 1.8-1.9~TB/s.
Finally, static unified variables with \verb|__managed__| have the lowest bandwidth at 103~GB/s.

The performance of \verb|hipMallocManaged| depends on whether XNACK is enabled. As shown in Table~\ref{tab:allocator}, with XNACK disabled, it allocates up-front, while with XNACK enabled, it allocates on-demand. Disabled XNACK gives higher bandwidth for both CPU and GPU.

On the CPU, the best achieved bandwidth was either 208~GB/s (case A) or around 180~GB/s (case B). The baseline bandwidth with \verb|malloc| memory is 181~GB/s, while the bandwidth with HIP allocators is 208~GB/s. If the memory is first touched by the GPU, then \verb|malloc| memory also achieves 208~GB/s. \verb|hipMallocManaged| with XNACK performs similarly to \verb|malloc| at 179~GB/s.

In case A, the peak bandwidth was reached with 24~threads (i.e. when all 24~cores were used). In contrast, in case B, the peak bandwidth was reached with only 9~threads, with performance dropping to 173-176~GB/s when using all cores (not pictured).
 
Regardless, the CPU is far from utilizing the full bandwidth of the memory with only 3\% of the theoretical peak, compared to 67\% for the GPU. For bandwidth-bound codes, the GPU has a clear advantage over the CPU. We found that the highest bandwidth is provided by \verb|hipMalloc|, which is 1.6--2.0~times faster than other options on the GPU. On the CPU, on-demand allocators have a disadvantage compared to up-front allocators, unless the data is GPU-initialized, in which case all allocators provide the same bandwidth.

\subsection{Legacy CPU--GPU Data Transfers}
Using \verb|hipMemcpy| between "host memory" (i.e. \verb|malloc| or \verb|hipHostMalloc|)) and "GPU memory" (i.e. \verb|hipMalloc|) is significantly slower than the achievable memory bandwidth. \verb|hipMemcpy| only achieves a peak bandwidth of 58~GB/s, or 850~GB/s when SDMA is disabled. However, "GPU to GPU" memory transfer (i.e. \verb|hipMalloc| to \verb|hipMalloc|) can reach close to the GPU memory bandwidth at 1900~GB/s. A possible explanation is that \verb|hipMemcpy| uses DMA transfers, which are more expensive when buffers are not page-locked (as in the case of \verb|malloc|).

\subsection{Coherence Overhead}

\begin{figure}
    \centering
    \includegraphics[width=\linewidth]{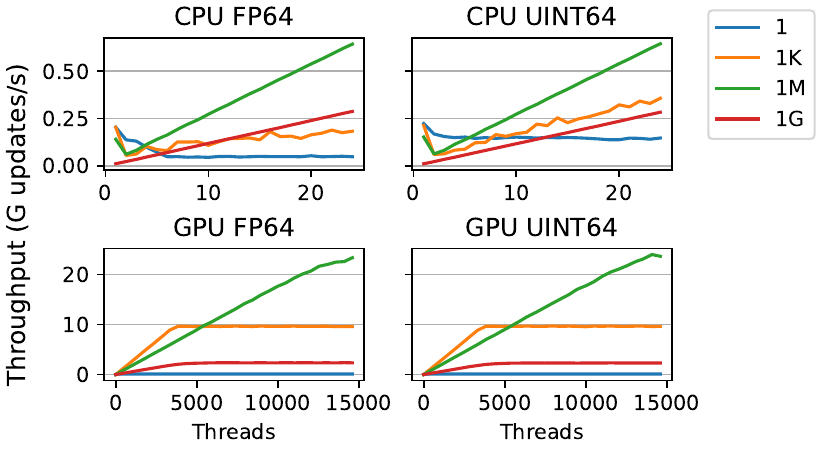}
    \caption{Atomics throughput in billion updates/s on an array with $2^0$, $2^{10}$, $2^{20}$, or $2^{30}$ elements. Note the different axis scales.}
    \label{fig:atomic_iso}
\end{figure}

\textbf{Isolated Performance.} The CPU results are shown in the first row of Fig.~\ref{fig:atomic_iso}. For the three smaller array sizes, the throughput at 1 thread is higher than 2 or 3 threads, due to the introduced coherence overhead. On the 1~M array, the 1 thread case is overtaken with 6 threads and continues to scale linearly. 1~G also scales linearly but with a lower slope. These large arrays scale well since collisions between threads are relatively unlikely. 1~M is faster as it fits inside L2 cache, while 1~G requires frequent accesses to main memory. The smaller arrays 1 and 1~K have more collisions between threads. With only 1 element, performance decreases with the number of threads. The integer version (UINT64) is about $3\times$ faster than the floating-point version (FP64). Interestingly, on the 1~K array, the FP64 version is similar or slower than 1~G, while the UINT64 version is consistently faster than 1~G.

The GPU results are shown in the second row of Fig.~\ref{fig:atomic_iso}. The GPU exhibits the same performance for the FP64 and UINT64 versions and is significantly higher than the CPU performance, except when using very few threads (1 or 64) or if there is only one element. Similar to the CPU, the 1~M case has the highest throughput and scales linearly with the number of threads.

The compiler generated native \verb|atomic_add| instructions for both integer and floating-point versions for the GPU. However, for the CPU, the compiler generated \verb|lock incq| (atomic increment) instructions for integers, but CAS loops (using \verb|lock cmpxchgq|) for floats because the x86 instruction set does not support native atomic floating-point operations. Collisions are more expensive with the CAS loop as they lead to extra iterations through the loop. Therefore the 1 and 1~K array sizes are slower with FP64 than UINT64 on the CPU.

\begin{figure}
    \centering
    \includegraphics[width=\linewidth]{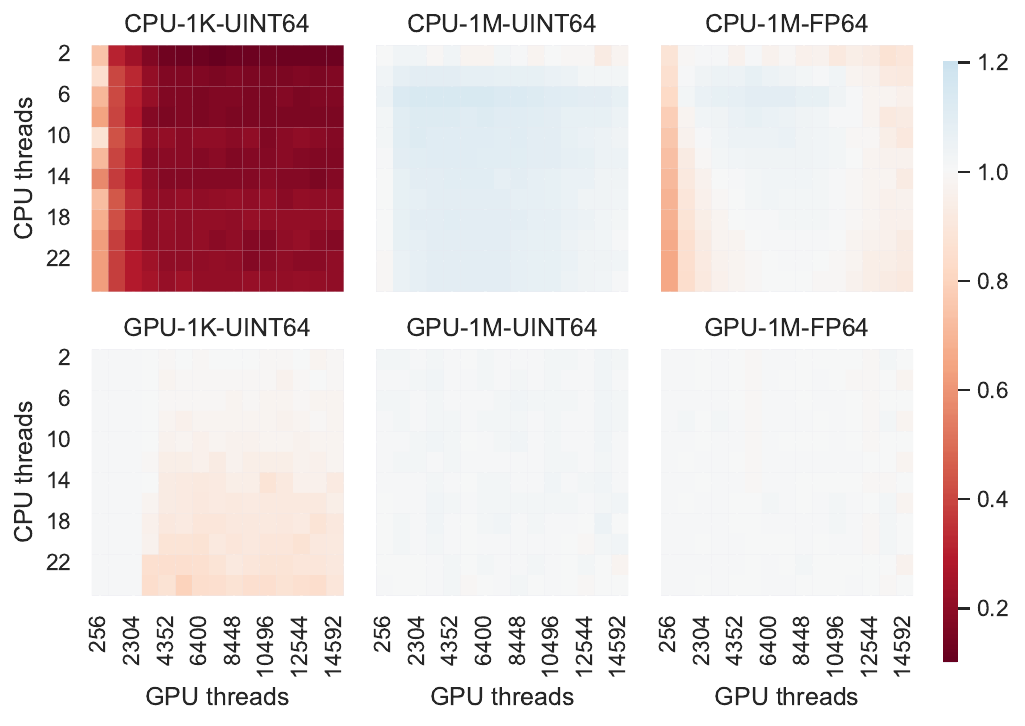}
    \caption{Relative performance of the CPU (first row) and GPU (second row) atomics performance when co-running.}
    \label{fig:atomic_hybrid}
\end{figure}

\textbf{Hybrid CPU--GPU Overhead.} Fig.~\ref{fig:atomic_hybrid} presents the relative performance of co-running CPU and GPU threads, as compared to the isolated performance in Fig.~\ref{fig:atomic_iso}.
The 1K array has the highest contention of the tested hybrid cases. This affects the CPU performance more than the GPU performance. The CPU performance is at best within 13\% of the baseline, but with 3328 GPU threads or more the relative CPU performance is only between 11\%--25\%. Below 3328 GPU threads, the GPU performance is similar to the baseline. With an increasing number of CPU and GPU threads, the GPU performance drops off to 79\%.

The 1M array has lower contention and thus higher performance. Counter-intuitively, with UINT64 the performance is slightly improved in most configurations compared with the isolated baseline. The CPU improvement is largest with 6 CPU threads and 2304--6400 GPU threads with a speedup of $1.14\times$. The GPU speedup is 1.02-1.03$\times$ in many cases, with a geometric mean of $1.01\times$. With FP64, the CPU performance is lower. There is still a region of speedup centered around the same thread configurations as for UINT64. However, with fewer than 1280 or more than 10496 GPU threads, the CPU kernel is slower than the baseline. The GPU performance is similar to the baseline with a geometric mean of 1.00.

In summary, atomics can be used to synchronize threads on both CPU and GPU. The GPU can perform more atomic operations per second than the CPU, while keys to improve atomics performance are minimizing the probability of collisions and ensuring that the dataset can fit in L2 cache. The CPU FP64 performance is even more sensitive to contention since it does not support native floating-point atomics. These effects also carry over to CPU--GPU hybrid algorithms, where the CPU is more disadvantaged than the GPU by contention. 

\section{System Software for Memory Management}
\label{sec:alloc}
In this section, we evaluate the effectiveness of memory management for memory allocation, page fault handling, TLB management, and Infinity Cache utilization. 

\subsection{Memory Allocation}
\begin{figure}[bt]
    \centering
    \includegraphics[width=\linewidth]{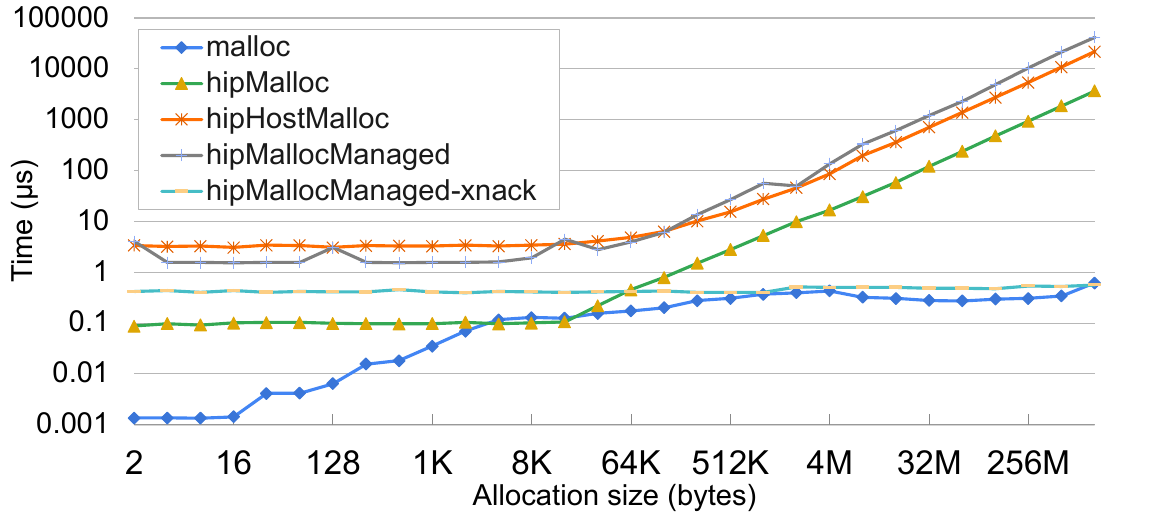}
    \caption{The memory allocation time ($\mu$s) using different allocators for increased allocation sizes (log--log plot).}
    \label{fig:alloc_time}
\end{figure}

The fastest allocator is \verb|malloc|, taking only 14~ns for allocating 32~B and 6~\si{\us} for  1~GiB, as shown in Figure~\ref{fig:alloc_time}. This is expected since \verb|malloc| is an on-demand allocator that does not allocate physical pages until first touch. The time for all up-front allocators is constant for allocating up to 16~KiB, indicating that this is their minimum granularity of physical memory allocation. The pattern is most revealing in \verb|hipMalloc|, which takes 10~\si{\us} up to 16~KiB, and then scales to 37~ms at 1~GiB. Finally, \verb|hipHostMalloc| and \verb|hipMallocManaged| (without XNACK) follow a similar curve, from around 15--34~\si{\us} up to 16~KiB and then scaling to 200--400~ms at 1~GiB. Note that \verb|hipMallocManaged| with XNACK enabled becomes an on-demand allocator, however, its execution time is constant regardless of allocation size. We believe it is caused by the overhead in the HIP implementation that is optimized for discrete GPUs.

The deallocation (figure omitted) follows a similar pattern. Interestingly, \verb|free| is faster than \verb|malloc| until 16~MiB. From 32~MiB, \verb|free| takes 4--9x longer time than \verb|malloc|. For \verb|hipMalloc|, deallocation is faster than allocation until 2~MiB, from which deallocation becomes significantly slower than allocation by up to 22x at 256~MiB. Freeing allocation by \verb|hipMallocManaged| with XNACK takes 3--21~\si{\us} while freeing \verb|hipHostMalloc| and \verb|hipMallocManaged| (no XNACK) memory takes from 220~\si{\us} to 67~ms at 1~GiB.

Overall, the recommended interface is \verb|malloc| for on-demand memory and \verb|hipMalloc| for up-front memory. For most allocations, \verb|malloc| provides the fastest allocation and de-allocation. However, on-demand allocators pay the page fault cost at runtime, which depends on how densely or sparsely the application touches the allocated memory (see Section~\ref{sec:fault}). \verb|hipMalloc| is the fastest up-front allocator and should be used for applications that want to avoid page fault cost at runtime.

\subsection{Page Fault Overhead}\label{sec:fault}

\begin{figure}[bt]
    \centering
    \includegraphics[width=\linewidth]{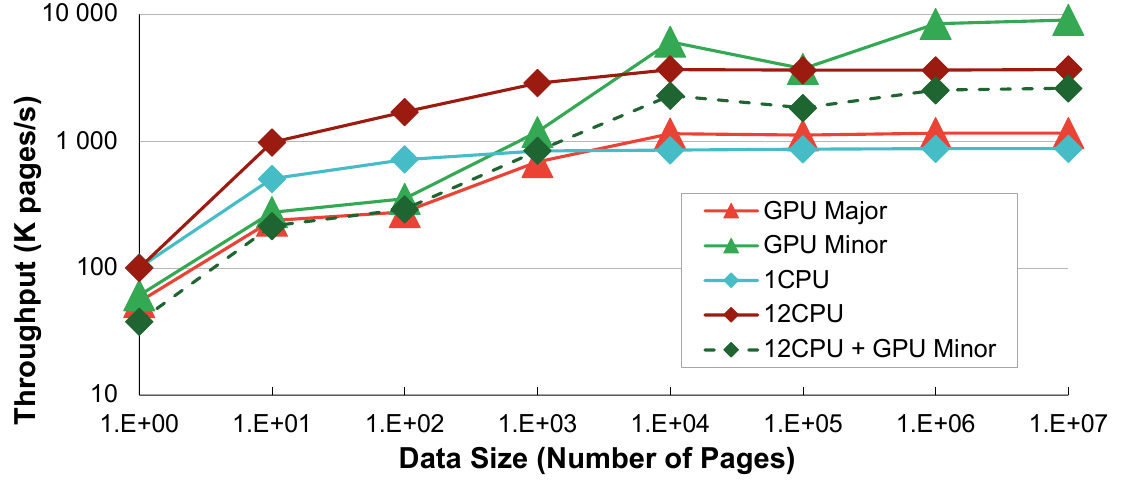}
    \caption{The throughput of page faults in pages/s in various scenarios: first-touch fault on GPU (GPU Major); first-touch on CPU and fault on GPU (GPU Minor); fault on one CPU core (1CPU); fault on 12 CPU cores (12CPU). Log--log plot.}
    \label{fig:fault_thruput}
\end{figure}

We evaluate the overhead of page faults on GPU and CPU in terms of throughput and latency. Throughput measures the maximum number of concurrent page faults that can be handled per second while latency measures the minimum time needed for handling a single page fault. 

The measured throughput of page fault handling is presented in Fig.~\ref{fig:fault_thruput}. It initially increases with the number of pages until reaching a plateau at $10^5$-$10^6$ pages. GPU Major reaches a steady state at 10~K pages with around 1.1~M~pages/s, while GPU Minor throughput increases up to 9.0~M~pages/s at 10~M pages, corresponding to a third of the total memory capacity. A single CPU core saturates at 1~K pages, reaching 872~K~pages/s while the 12-core CPU case saturates at 10~K pages, reaching 3.7~M~pages/s. The throughput of pre-faulting on CPU and then minor faulting on the GPU (12CPU + GPU Minor), compared to the GPU Major case, achieves up to $2.2\times$ improvement at 10~M pages (40~GiB).

\begin{figure}[bt]
    \centering
    \includegraphics[width=\linewidth]{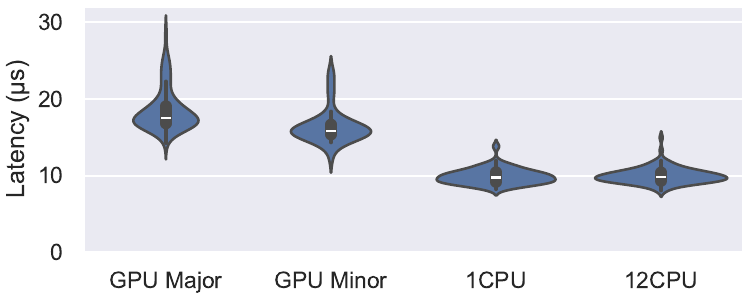}
    \caption{The distribution of latency for resolving a single page fault on GPU and CPU.}
    \label{fig:fault_latency}
\end{figure}

Faults on the CPU have lower latency than GPU faults, as shown in Figure~\ref{fig:fault_latency}. The CPU single-page latency is 9~\si{\us} on average with 11~\si{\us} tail latency (95\textsuperscript{th} percentile). The GPU latency is 1.8-2.0~times higher with 16~\si{\us} for a minor fault and 18~\si{\us} for a major fault. The GPU tail latency is also higher with 20~\si{\us} for minor and 22~\si{\us} for major faults, indicating higher variability.

The recommended strategy for applications exhibiting high concurrent page faults on GPU is to use CPU pre-faulting to transform them into GPU minor faults in advance. This is also an effective strategy for applications whose GPU runtime is dominated by GPU fault latency. However, if the time of CPU pre-faulting cannot be overlapped with GPU computation, directly faulting on GPU reduces the total latency.

\subsection{Adaptive Memory Fragments}
\begin{figure}[bt]
    \centering
    \includegraphics[width=\linewidth]{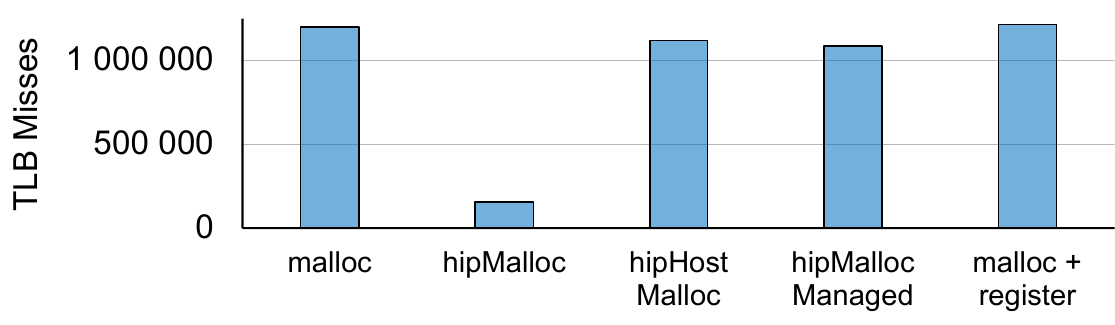}
    \caption{The number of measured GPU TLB misses in the TRIAD kernel using five allocators.}
    \label{fig:gpu_profile}
\end{figure}

The fragment size in the GPU page table is related to the number of TLB misses.
Fig.~\ref{fig:gpu_profile} presents the number of GPU TLB misses in STREAM. All configurations, except \verb|hipMalloc|, have 1.0-1.2~M TLB misses, while \verb|hipMalloc| has only 158~K misses. Since the driver sets the fragment field opportunistically based on contiguous pages, the number of GPU TLB misses on a memory range depends on the level of contiguity. On-demand allocators are naturally disadvantaged in this respect, as they allocate physical pages incrementally in a non-deterministic order. Up-front allocators can more easily ensure high contiguity by allocating all pages at once.

Our findings indicate that hipMalloc allocates memory with higher virtual and physical contiguity and thus uses larger fragment sizes in the GPU TLB. Consequently, memory from \verb|hipMalloc| has fewer TLB misses, explaining the significant bandwidth advantage of \verb|hipMalloc| shown in Section~\ref{sec:bw}.

\begin{figure}[bt]
    \centering
    \includegraphics[width=\linewidth]{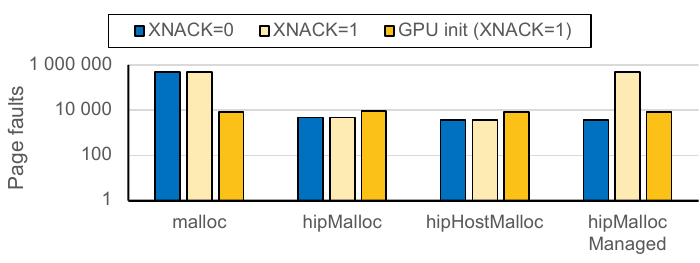}
    \caption{The total number of page faults in the CPU STREAM benchmark, 10 iterations (log scale). Three configurations: baseline (XNACK=0), XNACK=1, and GPU init (first-touch by GPU).}
    \label{fig:cpu_profile}
\end{figure}

\subsection{Infinity Cache Utilization}\label{sec:ic-util}
The CPU-side memory latency and bandwidth characterization in Section~\ref{sec:char} indicates that CPU-based allocators (i.e. \verb|malloc|) and CPU initialization may result in less effective utilization of the memory-side Infinity Cache, compared with HIP's up-front allocators (e.g. \verb|hipMalloc|) and GPU initialization. It cannot be explained by different fragment sizes since memory fragments are not used in the CPU page table. Also, all allocators lead to the same number of CPU-side TLB misses.

Instead, a possible explanation for the difference in cache effectiveness lies in the mapping of data to individual memory channels, as the Infinity Cache is partitioned into slices mapped to individual memory channels~\cite{cdna3-whitepaper}.
Physical pages are interleaved among the eight memory stacks at a 4~KiB granularity~\cite{cdna3-whitepaper}. The allocator must allocate the same number of physical pages from each corresponding physical range to evenly distribute data across channels. Any bias in the physical address mapping would result in less effective utilization of the Infinity Cache and thus higher latency and lower bandwidth for data sizes near the Infinity Cache capacity. The observed results suggest that the GPU-based allocators evenly allocate physical addresses (likely by allocating larger contiguous chunks), while CPU-initialized \verb|malloc|-based memory has a larger bias in physical memory mapping.

Indeed, the number of CPU-side page faults (Fig.~\ref{fig:cpu_profile}) varies significantly depending on the allocator. The most number of faults, around 472~K, occur with \verb|malloc| and \verb|hipMallocManaged| with XNACK, while \verb|hipMalloc| and \verb|hipHostMalloc| only have 3.7-4.6~K faults (when CPU initialized) or 8.0-8.9~K faults (when GPU initialized). The difference in page faults indicates that memory allocation granularity differs.

In summary, our findings advise developers to use up-front memory allocation (e.g. \verb|hipMalloc|) or first-touch data on the GPU to ensure optimal physical address mapping for maximizing the utilization of the Infinity Cache.

\section{HPC Applications on UPM}
\label{sec:app}

\begin{figure}[bt]
    \centering
        \includegraphics[width=\linewidth]{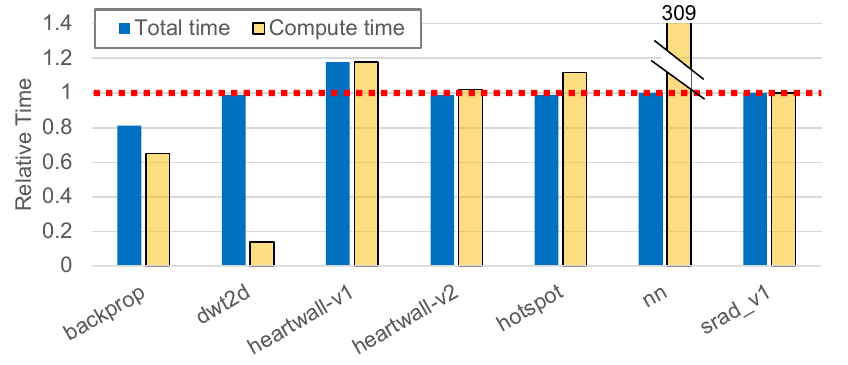}
        \includegraphics[width=\linewidth]{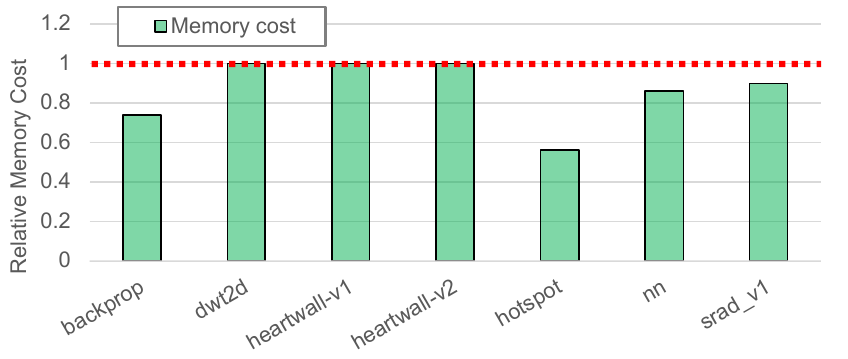}
    \caption{The performance of six applications using the UPM-enabled unified memory model, normalized to the baseline using the explicit model. Total execution time and compute time (upper plot) and memory usage (lower plot).}
    \label{fig:rodinia_results}
\end{figure}

We employed the strategies of Section~\ref{sec:strat} to port six HPC applications to the unified memory model. We summarize the implementation of each applied strategy as follows.
\begin{itemize}[leftmargin=*]
\item Concurrent accesses arise in heartwall due to the pipelining of pre-processing on the CPU with computing on the GPU. We used double buffering with stream events synchronization in the unified version.
\item In nn, hipGetMemInfo is used to calculate if the dataset will fit on the GPU. Our pragmatic solution was to remove the check and let the code fail if enough memory is not available.
\item Partial memory transfers are used in a pipeline to overlap data movement with computation in dwt2d and srad\_v1. In both cases, merging the buffers obviates the need for data movement altogether.
\item A scalar flag was stored as a stack variable in srad\_v1. The flag is set from a GPU kernel to determine the loop stop condition and is thus safe to access from the kernel.
\item Static memory is used extensively for both host and device data structures in heartwall. We created two versions: heartwall-v1 is close to the original code by using managed static variables, while heartwall-v2 is a restructured version without static variables and repeated allocation.
\item In nn, a \verb|std::vector| is initialized on the CPU and later passed to the GPU. We decided to keep the default vector in the unified version for simplicity.
\end{itemize}

Fig.~\ref{fig:rodinia_results} presents the relative memory usage, the total execution time, and the compute time of the unified memory version, compared to the baseline, for each application.
We use hipMalloc as the default unified allocator where possible, since it is the best performing in the characterization study.

The total execution time improved in the backprop application. By using the unified memory model, several data transfers in the main compute phase are removed, thereby reducing the compute time by 35\% and total time by 19\%. In dwt2d, the compute time was dominated by data transfer, and was reduced in the unified version by 86\%. However, the total execution time was dominated by I/O operations outside the compute phase, and thus the two versions result in similar total runtime. In srad\_v1, only a small amount of data transfer is performed in each iteration, while the runtime is dominated by kernel execution. Therefore, the compute time of srad\_v1 was not significantly affected.

Static managed variables are used in heartwall-v1, leading to an 18\% performance loss. In contrast, heartwall-v2 is a restructured version using dynamic memory allocation according to our porting strategy. This adaptation results in the unified memory model reaching the same performance as the explicitly managed version. 

There was one performance outlier in the form of the compute time in nn, which was significantly higher than the baseline. GPU page faults on the \verb|std::vector| significantly increased the compute time in the unified version. The magnitude of the increase is due to the relatively simple computational kernel compared to the cost of page faults. For optimal performance, the \verb|std::allocator| API could be applied to use \verb|hipMalloc| instead.

We also evaluate the memory usage of the applications. In four applications (backdrop, hotspot, nn, and srad\_v1), the peak memory usage was reduced by 10--44\% in the unified memory version, as their duplicated data in CPU and GPU buffers are merged into a unified buffer with UPM. In dwt2d and heartwall, the peak memory usage was unaffected in the unified version. In dwt2d, the peak memory usage occurs during the CPU-only IO phase, and is therefore not affected by unifying GPU data. In heartwall, the explicit host buffer plus device buffer have the same total memory usage as the UPM double buffering strategy.

In summary, the execution time of the unified memory version on UPM is competitive with the explicit model version. This is a significant step forward compared to UVM-based unified memory, which incurs high performance overhead for offering a simplified programming model~\cite{ganguly2019interplay,allen2021demystifying,allen2021depth}.
With similar performance, UPM-enabled unified memory programming further saves up to 44\% of memory usage in these applications, compared with the explicit model.

\section{Related Works}
Some previous works have studied UPM on MI300A in an application-specific context.
Tandon et al.~\cite{tandon2024porting} present OpenMP GPU offloading for UPM, porting the CFD code OpenFOAM to MI300A with OpenMP directives. Bertolli et al.~\cite{bertolli2024performance} identify a 1.2-1.3$\times$ improved performance in QMCPack with direct access to the APU GPU memory, compared to the copy configuration of discrete GPUs. They also report the potential overhead of page table initialization on the APU GPU and provide the configuration of eager maps as a solution. Markidis et al.~\cite{markidis2025exascale} ported the implicit particle-in-cell code iPIC3D to MI300A, observing only a 2\% overhead from using the unified memory model while enabling simulations with a larger number of particles on up to 32,768~APUs. Schieffer et al.~\cite{schieffer2025memsys} studied inter-APU communication on MI300A systems using micro-benchmarks and the proxy applications Quicksilver and CloverLeaf, finding that hipMalloc buffers provide the best communication performance. Nataraja et al.~\cite{nataraja2024enhanced} propose improvements to the system-level coherence for AMD APUs with a 14.4\% average performance improvement on a hardware simulator. Other works characterize the memory system of earlier architectures such as Grace Hopper with various memory allocation strategies, data placement, and memory access patterns~\cite{fusco2024understanding,schieffer2024harnessing}. 

Cooper et al.~\cite{cooper2024shared} investigate unified virtual memory in the Linux kernel’s HMM, and study the performance impacts on a diverse set of GPU workloads, revealing an aggressive prefetching strategy for demand paging. Landaverde et al.~\cite{landaverde2014investigation} investigate the performance of UVM in CUDA on synthetic and Rodinia benchmarks. They identify that UVM is limited by its high overhead and argue that the improvement in code complexity is not worthwhile. Chien et al.~\cite{chien2019performance} further examine the impact of memory prefetch and hints in CUDA UVM on application performance, showing the performance benefit from memory hints when the GPU memory is oversubscribed. In addition to UVM overhead and the impact on performance, Allen et al.~\cite{allen2021demystifying} analyze the effectiveness of prefetch and eviction techniques in fault elimination. Choi et al.~\cite{choi2022memory} focus on UVM for multi-GPU systems, providing a new approach to dynamically incorporate the spare memory of neighbor GPUs with a custom memory manager.

GPU profilers use a diverse set of methods, including API overloading, driver modification, and hardware event capture, to track the various memory behaviors like memory usage, page faults, etc. Lin et al.~\cite{lin2023drgpum} proposed the DrGPUM profiler to automatically identify inefficient memory coding practices in GPU-accelerated applications without modification to the application, hardware, or OS. They focused on problematic memory usage, object-level and intra-object memory inefficiencies, and GPU memory optimization targets for applications. Bachkaniwala et al.~\cite{bachkaniwala2024lotus} propose Lotus for profiling machine learning applications in PyTorch on GPUs. They focus on the preprocessing pipelines by linking the fine-grained timing of each preprocessing step to hardware-level events. Allen et al.~\cite{allen2021depth} modifies the GPU driver to track events associated with servicing on-demand faults in UVM. They focus on exploiting the batch features to mitigate the high overhead in UVM.

\section{Conclusions}
In summary, this work provides the first in-depth characterization of Unified Physical Memory for CPU and GPU in AMD MI300A, including the architectural properties of the memory system, the efficiency of system software for memory management, as well as application-level performance. In six HPC applications, UPM enables the unified memory model to achieve competitive performance compared to the explicitly managed model, while saving up to 44\% memory usage, when using our presented porting strategies. These results indicate that the unified memory model, once seen as a tradeoff of performance for programmability (due to software overhead in UVM), can now become the optimal choice on UPM for its high performance and significant memory saving.

\section*{Acknowledgments}
This work was performed under the auspices of the U.S. Department of Energy by Lawrence Livermore National Laboratory under Contract DE-AC52-07NA27344. LLNL-CONF-2004685. Funding from LLNL LDRD project 24-ERD-047 was used in this work. This research is supported by the Swedish Research Council (no. 2022.03062). 

\printbibliography

\end{document}